\documentclass[12pt]{article}
\usepackage{a4wide}
\usepackage{epsfig}
\usepackage{amssymb,amsmath}
\usepackage{setspace}
\pdfoutput=1
\usepackage{subfigure}
\usepackage{amssymb,amsmath}
\usepackage{graphicx}
\usepackage{color}
\usepackage{cancel}
\usepackage[colorlinks=true
,urlcolor=blue
,citecolor=blue
,linkcolor=blue
,pagecolor=blue
,linktocpage=true
,pdfproducer=medialab
]{hyperref}
\def\bi{\begin{itemize}}
\def\ei{\end{itemize}}
\newcommand{\Z}{{\mathbb Z}}

\def\to{\rightarrow}

\def\alt{\lesssim}
\def\agt{\gtrsim}

\newcommand\prd[3]{{Phys.\ Rev.\ }{\bf D #1} (#2) #3}
\newcommand\prl[3]{{Phys.\ Rev.\ Lett.\ }{\bf #1} (#2) #3}

\usepackage{amsmath}

\newcommand{\bea}{\begin{align}}
\newcommand{\eea}{\end{align}}

\newcommand{\beq}{\begin{equation}}
\newcommand{\eeq}{\end{equation}}

 \makeatletter
\def\alt{\mathrel{\mathpalette\gl@align<}}
\def\agt{\mathrel{\mathpalette\gl@align>}}
\def\gl@align#1#2{\lower.6ex\vbox{\baselineskip\z@skip\lineskip\z@
\ialign{$\m@th#1\hfil##\hfil$\crcr#2\crcr\sim\crcr}}} \makeatother

\usepackage{longtable}

\begin{document}
%\begin{flushright}
%BA-12-XX \\
%\end{flushright}
%
%\vspace*{0.5cm}

\begin{center}
{
\bf\LARGE
Naturalness and Dark Matter in \\[4mm]
a Realistic Intersecting D6-Brane Model }
\\[10mm]
{\large Waqas~Ahmed$^{\,\star}$} \footnote{E-mail: \texttt{waqasmit@itp.ac.cn}},
{\large Lorenzo~Calibbi$^{\,\star}$} \footnote{E-mail: \texttt{calibbi@itp.ac.cn}},
{\large Tianjun~Li$^{\,\star\,\heartsuit \,\dagger}$} \footnote{E-mail: \texttt{tli@itp.ac.cn}},\\[1mm]
{\large Shabbar~Raza$^{\,\ast}$} \footnote{E-mail: \texttt{shabbar.raza@fuuast.edu.pk}},
{\large Jia-Shu Niu$^{\,\star\,\heartsuit}$} \footnote{E-mail: \texttt{jsniu@itp.ac.cn }}
{\large Xiao-Chuan Wang$^{\,\diamond,}$}
\footnote{E-mail: \texttt{xcwang@live.com}}
%\end{center}
\\[10mm]
\centerline{$^{\star}$ \it
CAS Key Laboratory of Theoretical Physics,}
\centerline{\it
Institute of Theoretical Physics, Chinese Academy of Sciences,}
\centerline{\it Beijing 100190, P.\ R.\ China}
\vspace*{0.2cm}
\centerline{ $^\heartsuit $\it 
School of Physical Sciences, University of Chinese Academy
of Sciences,} 
\centerline{\it  Beijing 100049, P.\ R.\ China}
\vspace*{0.2cm}
\centerline{$^{\dagger}$ \it
School of Physical Electronics, University of Electronic Science and Technology of China,}
\centerline{\it Chengdu 610054, P.\ R.\ China}
\vspace*{0.2cm}
\centerline{$^{\ast}$ \it
Department of Physics, Federal Urdu University of Arts, Science and Technology,}
\centerline{\it Karachi 75300, Pakistan}
\vspace*{0.2cm}
\centerline{$^{\diamond}$ \it Department of Physics, Henan Normal University, Xinxiang, Henan, 453007, P.\ R.\ China}
\vspace*{1.cm}

%\vspace{1.cm} 
{\bf Abstract}
\end{center}
We revisit a three-family Pati-Salam model with a realistic phenomenology from intersecting D6-branes 
in Type IIA string theory compactified on a $\mathbf{T}^6/(\mathbb{Z}_2 \times \mathbb{Z}_2)$ orientifold, 
and study its naturalness in view of the current LHC and dark matter searches. We discuss spectrum
and phenomenological features of this scenario demanding fine tuning better than 1\%. 
This requirement restricts the lightest neutralino to have mass less than about 600 GeV. 
We observe that the viable parameter space is tightly constrained by the requirements of naturalness and consistency with 
the observed dark matter relic density, so that it is fully testable at  current and future  dark matter searches, unless 
a non-thermal production mechanism of dark matter is at work. 
We find that $Z$-resonance, $h$-resonance, $A$-funnel 
and light stau/stop-neutralino coannihilation solutions are consistent with current LHC and 
dark matter constraints while the ``well-tempered'' neutralino scenario is ruled out in our model. Moreover, 
we observe that only Bino, Higgsinos, right-handed staus and stops can have mass below 1 TeV. 
%We also present seven benchmark points corresponding to  $Z$-resonance, $h$-resonance, $A$-funnel and light stau/stop-neutralino coannihilation scenario, and comment on the Bino-Higgsino and Higgsino-type neutralino solutions as well.

\thispagestyle{empty}

%\bigskip
\newpage

%\addtocounter{page}{-1}
\addtocounter{footnote}{-6}

%%%%%%%%%%%%%%%%%%%%%%%%%%
%\baselineskip 30pt
% Main body
%%%%%%%%%%%%%%%%%%%%%%%%%%
\baselineskip 18pt
%%%%%%%%%%%%%%%%%%%%%%%%%%

\section{Introduction}

Despite the extensive searches performed at the Large Hadron Collider (LHC), 
no evidence for physics beyond the Standard Model (SM) has been found so far. 
Together with the observation of the Higgs boson, whose properties are within the uncertainties in good agreement with the SM predictions, this challenges the extensions of the SM that have been proposed to provide a natural explanation of the hierarchy between the electroweak symmetry breaking (EWSB) scale and the Planck scale. 
In particular, the limits set by the searches performed by the ATLAS and CMS collaborations on the mass of possible supersymmetric partners of the SM particles and the measured mass of the Higgs boson, $m_H \simeq 125$ GeV,
push low-energy supersymmetry (SUSY) -- once the most popular attempt to solve the gauge hierarchy problem -- 
into the range of fine tuning worse than the percent level, at least in the simplest SUSY-breaking scenarios. Therefore, we think that,
before giving up naturalness as a motivation for new physics, it is worth to survey possible exceptions to the above conclusion in the attempt of finding non-minimal and comparatively natural solutions. Indeed, several examples of such kind have been discussed in recent literature~\cite{Baer:2017pba,Calibbi:2016qwt,Ahmed:2016lkh}. In particular, an interesting scenario has been recently proposed, which was called `Super-Natural' SUSY~\cite{Leggett:2014hha, Du:2015una}. In this framework, no residual electroweak fine-tuning is left in the 
Minimal Supersymmetric Standard Model (MSSM) in presence of no-scale supergravity boundary conditions~\cite{Cremmer:1983bf} and Giudice-Masiero (GM) mechanism~\cite{Giudice:1988yz}, despite a relatively heavy spectrum.\footnote{Nevertheless, one might argue that the Super-Natural SUSY has a problem related to the higgsino mass parameter $\mu$, which is generated by the GM mechanism and is proportional to the universal gaugino mass $M_{1/2}$, since the ratio $M_{1/2}/\mu$ is of order one but cannot be determined as an exact number.  This problem, if it is,  can be addressed in a M-theory inspired Next to 
MSSM (NMSSM) \cite{Li:2015dil}.}

Apart from the gauge hierarchy problem, the most compelling motivation for new physics at energies accessible 
at the LHC is probably given by the possibility of explaining the observed Dark Matter (DM) in terms of 
a relic particle produced in the early Universe through the thermal freeze-out mechanism.
In Supersymmetric SMs (SSMs) with conserved $R$-parity, the Lightest Supersymmetric Particle (LSP) --
such as the lightest neutralino, the gravitino, etc.~-- is stable and can be a dark matter candidate. However,
the SSMs have in turn to fulfil the non-trivial constraints set by the DM abundance obtained from observations of 
the Cosmic Microwave Background (CMB). Furthermore, DM candidates have to face increasingly relevant 
constraints from DM searches, in particular direct detection experiments.

Another starting point of our work is the observation that string theory is one of the most promising candidates for quantum gravity.
Therefore, the goal of string phenomenology is to construct the SM or SSMs from string theory 
with moduli stabilization and without chiral exotics, and try to make unique predictions 
which can probed at the LHC and other future experiments.
In this article, we shall consider naturalness and dark matter phenomenology within intersecting D-brane models~\cite{Berkooz:1996km,
Ibanez:2001nd, Blumenhagen:2001te, CSU, Cvetic:2002pj, Cvetic:2004ui, Cvetic:2004nk, 
Cvetic:2005bn, Chen:2005ab, Chen:2005mj, Blumenhagen:2005mu},
where realistic SM fermion Yukawa couplings can be realized only within the Pati-Salam gauge group \cite{PS}. 
Three-family Pati-Salam models have been constructed systematically in
Type IIA string theory on the $\mathbf{T^6/(\Z_2\times \Z_2)}$
orientifold with intersecting D6-branes~\cite{Cvetic:2004ui}, and 
it was found that one model has a realistic phenomenology: the tree-level 
gauge coupling unification is achieved naturally around the string scale,  
the Pati-Salam gauge symmetry can be broken down to the SM close to the string
scale, the small number of extra chiral exotic states can be decoupled
via the Higgs mechanism and strong dynamics, the SM fermion masses and mixing
can be accounted for, the low-energy sparticle spectra may potentially
be tested at the LHC, and the observed dark matter relic density may be 
generated for the lightest neutralino as the LSP, and so on~\cite{Chen:2007px, Chen:2007zu,Li:2014xqa}. 
In short, this is one of the best globally consistent string models,
and represents one of the few concrete string models that is phenomenologically viable 
from the string scale to the EWSB scale, where it features the usual spectrum of 
the MSSM.

The aim of the present work is to assess the naturalness of the above-mentioned D-brane model in view of 
the LHC and DM constraints, and highlight spectra and other phenomenological features of 
the viable parameter space selected by requiring low fine-tuning. 
We base our naturalness considerations on a quantity called  `electro-weak' fine-tuning measure ($\Delta_{\rm EW}$) 
defined as \cite{Baer:2012up,Baer:2012mv}
\begin{align}
\Delta_{\rm EW} \equiv \frac{\max_a |C_a|}{m^2_Z/2},
\label{eq:DeltaEW}
\end{align}
where $C_a$ are the terms appearing in the right-hand side of the expression
\begin{align}
\frac{m_Z^2}{2} = \frac{(\widetilde{m}^2_{H_d} +\Sigma_d) - (\widetilde{m}^2_{H_u} +\Sigma_u)\tan^2\beta}{\tan^2\beta-1} - |\mu|^2,
\label{eq:ewsb}
\end{align}
which follows from minimization of the scalar potential. 
Here, $\widetilde{m}^2_{H_{u}}$ and $\widetilde{m}^2_{H_d}$ are the SUSY breaking soft mass terms 
of the two Higgs doublets, and $\tan\beta$ the ratio of their vacuum expectation values (vevs), while $\mu$ is the Higgs bilinear coupling appearing in the superpotential.
Explicit expressions for the quantities $\Sigma_{u,d}$, which encode 1-loop corrections to the tree-level potential, can be found in~\cite{Baer:2012cf}. 
All quantities in Eq.~(\ref{eq:ewsb}) are defined at low energy.
For moderate to large values of $\tan\beta$, the dominant contributions to $\Delta_{\rm EW}$ stem from $\widetilde{m}^2_{H_u}$ and $\mu^2$. In fact, it is typically a cancellation between these two terms that ensures the correct $Z$ mass in presence of heavy superpartners (stops and gluinos, in particular), 
whose effect is a radiative enhancement of  $|\widetilde{m}^2_{H_u}|$.

Based on what we found in previous works \cite{Calibbi:2016qwt,Ahmed:2016lkh}, we expect to find solutions with reduced fine tuning (FT) if the Wino mass is substantially larger than the gluino mass at the unification scale. In fact, this triggers a compensation between gauge and Yukawa radiative corrections to $\widetilde{m}^2_{H_u}$, reducing its sensitivity to stop and gluino masses. 
This effect can be spotted from $\beta$-function of $\widetilde{m}^2_{H_u}$, which at one loop is given by
\begin{align}
16\pi^2 \frac{d}{dt} 
\widetilde{m}^2_{H_u} \approx 6 y_t^2 \left[\widetilde{m}^2_{H_u} + \widetilde{m}^2_{Q_3}+ \widetilde{m}^2_{U_3}+A_t^2 \right]  - 6g_2^2M_2^2, 
\label{eq:rge}
\end{align}
where the hypercharge-dependent terms were omitted. 
The term controlled by the top Yukawa $y_t$ 
(there, $\widetilde{m}^2_{Q_3}$ and $\widetilde{m}^2_{U_3}$ are the left-handed and right-handed stop masses respectively, 
and $A_t$ the stop trilinear term) carry an opposite sign with respect to the $SU(2)$ gauge term proportional to the Wino mass $M_2$, such
that a compensation between the two terms,  hence a reduced low-energy value of $|\widetilde{m}^2_{H_u}|$, is possible provided that $M_2>M_3$ (given that the gluino mass $M_3$ induces large positive contributions to the stop masses in the running). As we will see in the next section, this kind of non-universality of the gaugino mass terms can be easily achieved in our D-brane model, so that it will be a feature of the regions of the parameter space selected by requiring low values of $\Delta_{\rm EW}$.

The rest of the paper is organized as follows. In Section \ref{sec:model}, we review 
the features of the model that are relevant for our study. 
We describe how we preform the parameter space scan and which phenomenological constraints we impose
 in Section \ref{sec:scan}. We present our numerical results in Section \ref{sec:numerics}, 
and in \ref{sec:conclusions} we summarize and conclude.

%%%%%%%%%%%%%%%%%%%%%%%%%%%%%%%%%%%%%%%%%%%%%%%%%%%%%%%%%%%%%%%%
\section{The Realistic Pati-Salam Model from Intersecting D6-Branes}
%%%%%%%%%%%%%%%%%%%%%%%%%%%%%%%%%%%%%%%%%%%%%%%%%%%%%%%%%%%%%%%%
\label{sec:model}

We are going to study the realistic intersecting D6-brane model proposed in Ref.~\cite{Cvetic:2004ui}, based 
on Type IIA string theory compactified on a $\mathbf{T}^6/(\mathbb{Z}_2 \times \mathbb{Z}_2)$ orientifold, whose
 appealing phenomenological features have been briefly reviewed in the Introduction.
Supersymmetry is broken by the F-terms of the dilaton $S$ and three complex structure moduli $U_i$, 
respectively $F^S$ and $F^{U_i}$, $i=1,3$.
Neglecting the CP-violating phases, the resulting soft terms can be parametrized
by the gravitino mass $m_{3/2}$, and the angles $\Theta_1$, $\Theta_2$, $\Theta_3$
for the complex structure moduli directions, and $\Theta_4\equiv \Theta_s$ for the dilaton one, 
which are related by~\cite{Chen:2007zu}
\begin{align}
\sum_{i=1}^4\Theta^2_i=1.
\end{align}
In terms of these parameters, the soft SUSY-breaking terms at the Grand Unification (GUT) scale can be written as~\cite{Chen:2007zu} 
\begin{align}
& M_1 =(0.519\Theta_1+0.346\Theta_{2}+0.866\Theta_{3})\times m_{3/2} ~,~\nonumber \\
& M_2 =(0.866\Theta_2 - 0.866\Theta_{4} )\times m_{3/2} ~,~\nonumber \\
& M_3 =(0.866\Theta_2 + 0.866\Theta_{3})\times m_{3/2} ~,~\nonumber \\
& A_0 =(-1.111\Theta_{1}-0.621\Theta_{2}+0.245\Theta_{3}-0.245\Theta_{4})\times m_{3/2} ~,~\nonumber \\
& \widetilde{m}_L =\sqrt{1.0+0.899\Theta_{1}^{2}-0.518\Theta_{2}^{2}-0.849\Theta_{3}^{2}-1.418\Theta_{4}^{2}-0.557\Theta_{1}\Theta_2-0.557\Theta_{3}\Theta_{4}}\times m_{3/2} ~,~\nonumber \\
& \widetilde{m}_R =\sqrt{1.0-1.418\Theta_{1}^{2}-0.849\Theta_{2}^{2}-0.518\Theta_{3}^{2}+0.899\Theta_{4}^{2}-0.557\Theta_{1}\Theta_{2}-0.557\Theta_{3}\Theta_{4}}\times m_{3/2} ~,~\nonumber\\
& \widetilde{m}_{H_u}=\widetilde{m}_{H_d}=\sqrt{1.0-1.5 \Theta_{3}^2-1.5 \Theta_{4}^2}\times m_{3/2}~, 
\label{ssb}
\end{align}
where $M_{1,2,3}$ are the gauginos masses, $A_0$ is a common trilinear term, and $\widetilde{m}_L$ and $\widetilde{m}_R$
are the soft mass terms for, respectively, the left-handed and right-handed squarks and sleptons. Notice the Pati-Salam-symmetric structure of
the soft terms.

%%%%%%%%%%%%%%%%%%%%%%%%%%%%%%%%%%%%%%%%%%%%%%%%%%%%%%%%%%%%%%%%
\section{Scanning Procedure and Constraints}
%%%%%%%%%%%%%%%%%%%%%%%%%%%%%%%%%%%%%%%%%%%%%%%%%%%%%%%%%%%%%%%%
\label{sec:scan}
We employ the ISAJET~7.85 package~\cite{ISAJET} to perform random scans over the parameter space of the D-brane model presented in the previous section.
Following \cite{Li:2014xqa}, we rewrite the three independent $\Theta_i$ parameters that enter the soft masses in (\ref{ssb}) as
\begin{align}
&\Theta_1 = \cos(\beta)\cos(\alpha)\sqrt{1-\Theta_{4}^{2}} ,~\nonumber\\
&\Theta_2 = \cos(\beta)\sin(\alpha)\sqrt{1-\Theta_{4}^{2}} ,~\nonumber \\
&\Theta_3 = \sin(\beta)\sqrt{1-\Theta_{4}^{2}} ,~\nonumber\\
&{\rm where}\quad \alpha  \equiv 2 \pi \gamma_{1} ,~ \beta \equiv 2 \pi \gamma_{2} .
\end{align}
We employ the Metropolis-Hastings algorithm described in \cite{Belanger:2009ti}
to scan over the following ranges of our parameters:
\begin{align}
0\leq & \gamma_1  \leq 1  ~,~\nonumber \\
0\leq & \gamma_2  \leq 1 ~,~\nonumber \\
0\leq &  \Theta_4  \leq 1 ~,~\nonumber \\
0\leq & m_{3/2}  \leq 11 ~ \rm{TeV} ~,~\nonumber \\
2\leq & \tan\beta  \leq 60~,
 \label{input_param_range}
\end{align}
For what concerns the SM parameters ({\it e.g.}~the top and bottom masses), we keep the values coded into ISAJET.

We only collect data points that satisfy the requirement of a successful radiative EWSB (REWSB), 
{\it i.e.}~a valid solution of Eq.~(\ref{eq:ewsb}), and
choose $\mu>0$. We also select the points with the lightest neutralino as the LSP.
Furthermore, we consider the following constraints that we apply as specified in the next section.
\\[2mm]
\textbf{LEP constraints.}
We impose the bounds that the LEP2 experiments set on charged sparticle masses ($\gtrsim 100$ GeV) \cite{Patrignani:2016xqp}.
\\[2mm]
\noindent
\textbf{Higgs mass.}
The experimental combination for the Higgs mass reported by 
the ATLAS and CMS Collaborations is \cite{Khachatryan:2016vau}
\begin{align}\label{eqn:mh}
m_{h} = 125.09 \pm 0.21(\rm stat.) \pm 0.11(\rm syst.)~GeV .
\end{align}  
Due to an estimated 2 GeV theoretical uncertainty in the calculation of $m_h$ in the MSSM -- see {\it e.g.}~\cite{Allanach:2004rh} --
we consider the following range 
\begin{align}\label{eqn:higgsMassLHC}
123~ {\rm GeV} \leq m_h \leq 127~ {\rm GeV}. 
\end{align}
\noindent
\textbf{$B$-physics constraints.}
We use the IsaTools package \cite{bsg,bmm} to compute the following observables and set the $2\sigma$ constraints:
\begin{align}\label{eqn:Bphysics}
1.6\times 10^{-9} \leq ~ {\rm BR}(B_s \rightarrow \mu^+ \mu^-) ~
  \leq 4.2 \times10^{-9} \quad \cite{CMS:2014xfa},\\ 
2.99 \times 10^{-4} \leq  ~ {\rm BR}(b \rightarrow s \gamma) ~
  \leq 3.87 \times 10^{-4} \quad \cite{Amhis:2014hma},\\
0.70\times 10^{-4} \leq ~ {\rm BR}(B_u\rightarrow\tau \nu_{\tau})~
        \leq 1.5 \times 10^{-4} \quad \cite{Amhis:2014hma}.
\end{align}
\noindent
\textbf{Electroweak fine tuning.}
As discussed in the Introduction, we are interested in focusing on comparably natural scenarios. 
Therefore, we are going to consider regions of the parameter space with a tuning better than the $\sim1\%$, 
{\it i.e.}~for which the electroweak fine-tuning measure defined in Eq.~(\ref{eq:DeltaEW}) satisfies
\begin{align}\label{eqn:deltaEW}
\Delta_{\rm EW} < 100.
\end{align}
\noindent
\textbf{LHC searches.}
Instead of a full recasting of the overwhelming number of searches for SUSY particles performed 
by the LHC Collaborations, here we only employ 
the latest analyses interpreted in terms of simplified models, in order to obtain approximate limits on the spectra. 
This approach is justified by the relative simplicity of the spectrum of our scenario, in particular for what concerns the possible light particles, such that the simplified models represent a reasonable approximation. In fact, 
our FT requirement in Eq.~(\ref{eqn:deltaEW}) is achieved for rather heavy Winos, as explained below Eq.~(\ref{eq:rge}), which radiatively increases the masses of all particles charged under $SU(2)_L$. Furthermore, the condition (\ref{eqn:deltaEW})
requires relatively light Higgsinos and thus neutralino LSP ($\mu \lesssim 600$ GeV, see {\it e.g.}~\cite{Calibbi:2016qwt,Ahmed:2016lkh}). This implies
that the limits on gluinos and the first/second generation squarks from searches based on multi-jets and missing energy  are very robust, and, due to the unified relations for the scalar masses in (\ref{ssb}), affect sleptons too. In the end, only Bino, Higgsinos, right-handed stau and stop are possibly light.
Based on \cite{ATLAS:2017cjl,Sirunyan:2017cwe,Sirunyan:2017kqq}, we consider the following condition on gluino and first/second generation squark masses
\begin{align}
(a) \quad \quad m_{\widetilde g} > ~ 2 ~ {\rm TeV},\quad \quad m_{\widetilde q} > ~ 2 ~ {\rm TeV},
\label{lhc-a}
\end{align}
which follows from the fact that the LSP is way below 1 TeV in the scenario under consideration, and we have $m_{\widetilde q}\sim m_{\widetilde g}$ as a consequence of both the boundary conditions in (\ref{ssb}) and the gluino radiative effects to squark mass terms.\\ \noindent
Searches for two and three leptons plus missing energy \cite{CMS:2017fdz,ATLAS:2017uun} set bounds on the electro-weak production of charged-neutral Higgsinos decaying to $WZ$ and the LSP, which we can approximately translate (cf.~\cite{Ahmed:2017jxl}) into the following condition 
% \textcolor{red}{[In our last paper, points with Higgsino dominant $\tilde\chi^{0}_{2} / \tilde\chi^{\pm}_{1}$ mass 
% up to 300 GeV are excluded in $Z$-pole scenario while for Higgs-pole scenario, the points with $\tilde\chi^{0}_{2}$ 
% mass up to 460 GeV are excluded. Since we have bino LSP for neutralino less than 100 GeV, I think this bound is ok.]}
\begin{align}
(b) \quad \quad {\rm if} ~~ m_{\widetilde \chi^{0}_{1}} < ~ 100 ~ {\rm GeV} ~\Longrightarrow ~ \mu > 350 ~ {\rm GeV}.
\label{lhc-b}
\end{align}
Finally, searches for stops \cite{Sirunyan:2017cwe,Sirunyan:2017kqq,CMS:2017vbf,ATLAS:2017kyf,Aaboud:2016tnv}, including the compressed mass region, conservatively approximate to
\begin{align}
(c)&  \quad \quad {\rm if} ~~ m_{\widetilde \chi^{0}_{1}} < ~ 400 ~ {\rm GeV}~~ {\rm and} ~~(m_{\widetilde t_{1}}-m_{\widetilde \chi^{0}_{1}}) > 100 ~ {\rm GeV}~ \Longrightarrow ~m_{\widetilde t_{1}} > ~ 1 ~ {\rm TeV}, \label{lhc-c} \\
(d)& \quad \quad {\rm if} ~~ 10 < (m_{\widetilde t_{1}}-m_{\widetilde \chi^{0}_{1}}) < ~ 100 ~ {\rm GeV} ~\Longrightarrow ~m_{\widetilde t_{1}} > ~ 500 ~ {\rm GeV},
\label{lhc-d}\\
(e)& \quad \quad {\rm if} ~~  (m_{\widetilde t_{1}}-m_{\widetilde \chi^{0}_{1}}) < ~ 5 ~ {\rm GeV} ~\Longrightarrow ~m_{\widetilde t_{1}} > ~ 323 ~ {\rm GeV}.
\label{lhc-e}
\end{align}
\noindent
\textbf{DM searches and relic density.}
For the discussion on the phenomenology of neutralino DM in our scenario, we consider the following conservative range for the neutralino relic density, 
based on the results of \cite{planck2015}:
\begin{align}\label{eq:omega}
0.09 \, \le \Omega_\chi h^2 \le 0.14.
\end{align}
We are also going to show the impact of direct searches for DM considering the limits on the spin-independent (SI) and spin-dependent (SD) DM cross section with nuclei as presented in \cite{Akerib:2016vxi,Aprile:2017iyp,Aprile:2015uzo,Akerib:2017kat,Akerib:2016lao}.

%%%%%%%%%%%%%%%%%%%%%%%%%%%%%%%%%%%%%%%%%%%%%%%%%%%%%%%%%%%%%%%%
\section{Results and Discussion}
%%%%%%%%%%%%%%%%%%%%%%%%%%%%%%%%%%%%%%%%%%%%%%%%%%%%%%%%%%%%%%%%
\label{sec:numerics}
%
%%%%%%%%%%%%%%%%%%%%%%%%%%%%%%%%%%%%%%
\begin{figure}[t!]
\centering
\includegraphics[width=0.5\textwidth]{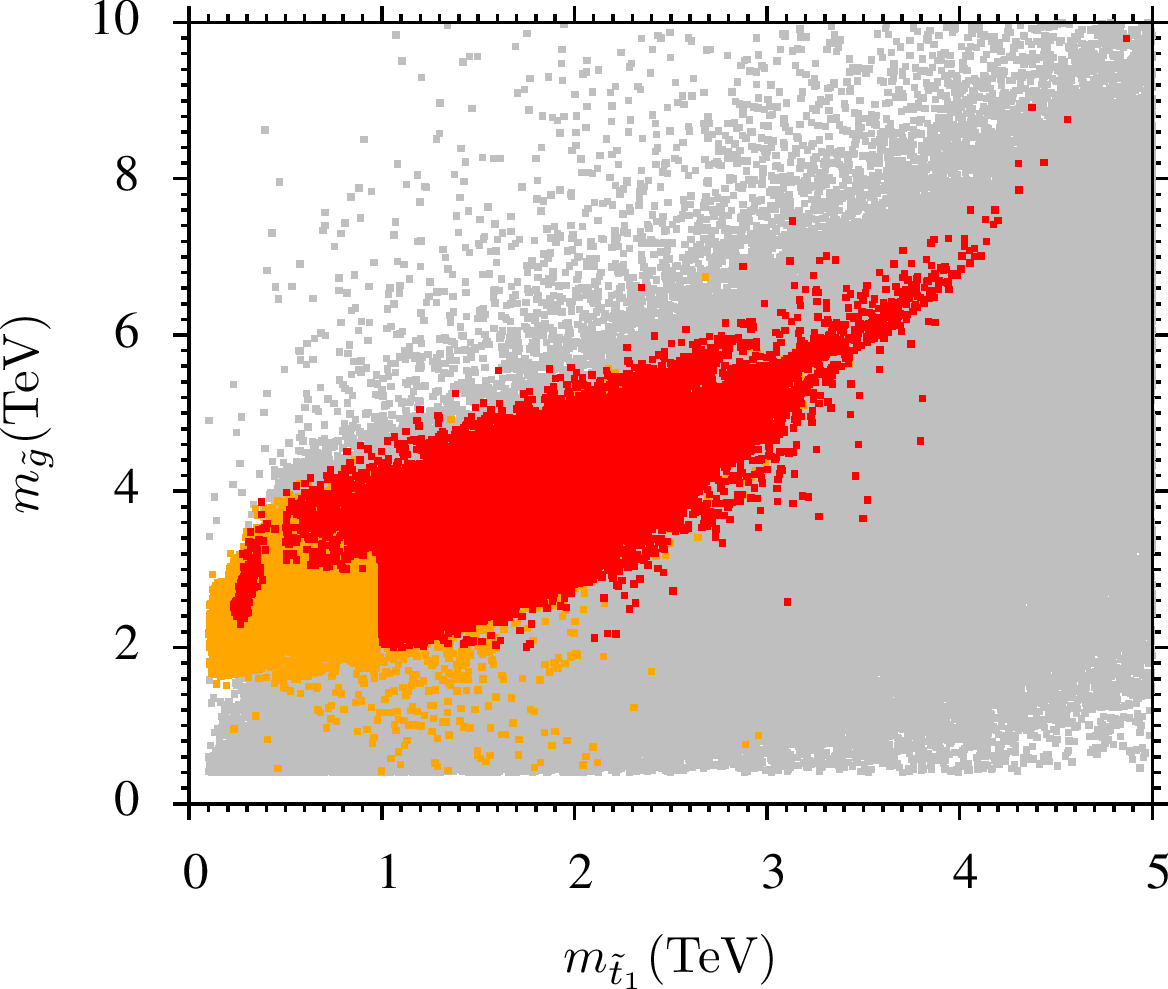}
\caption{\small Result of our scan displayed on the lightest stop-gluino mass plane. All points fulfil our `naturalness condition', $\Delta_{EW}\leq$100.
 Grey points satisfy the REWSB, yield a neutralino LSP and are consistent with LEP bounds.  
 Orange points give in addition a Higgs mass in the range (\ref{eqn:higgsMassLHC}), while 
 red points also satisfy $B$-physics and the LHC bounds described in Section~\ref{sec:scan}.} 
 \label{fig1}
\end{figure}
%%%%%%%%%%%%%%%%%%%%%%%%%%%%%%%%%%%%%
%
As explained above, we focus on regions of the parameter space of our D-brane scenario that corresponds to $\Delta_{\rm EW} \leq 100$, {\it i.e.}~are still able to provide a relatively natural solution to the hierarchy problem. In Fig.~\ref{fig1}, we show the resulting points on the plane of the lightest stop mass vs.~the gluino mass. The grey points in the background fulfil the basic constraints discussed in the previous section. The orange points also give the correct Higgs mass,
and the red ones satisfy in addition the constraints from $B$-physics observables and our approximate LHC exclusion limits, Eqs.~(\ref{lhc-a}\,-\ref{lhc-e}).
This clearly shows that the LHC searches for production of strongly-interacting SUSY partners have the capability to test in part our parameter space with low tuning and have in fact excluded a corner of it already. This is in contrast to the case of models where the condition $M_2>M_3$ that reduce the sensitivity of 
$\widetilde{m}^2_{H_u} $ on stop and gluino masses (cf.~Eq.~(\ref{eq:rge}) and the discussion below it) is purely achieved by non-universal gaugino masses 
in gauge mediation \cite{Calibbi:2016qwt}. In fact, the spectra of such models are
 generally beyond the reach of the LHC. 

\begin{figure}[t!]
\centering
\includegraphics[width=0.5\textwidth]{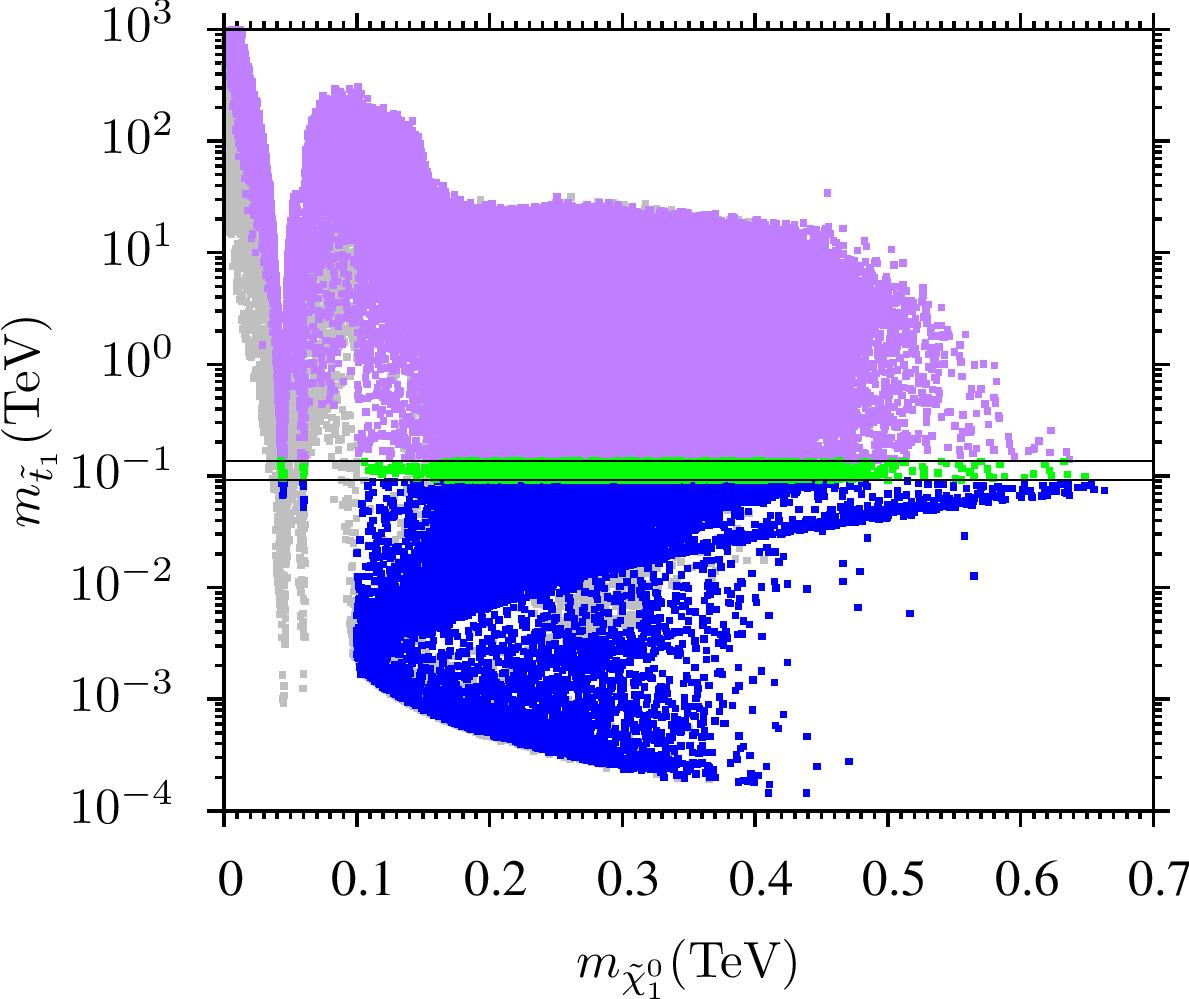}
\caption{\small Neutralino relic density $\Omega_\chi h^{2}$ vs.~its mass, $m_{\widetilde \chi_{1}^{0}}$. All points fulfil $\Delta_{EW}\leq$ 100.
Grey points satisfy all the constraints discussed in Section \ref{sec:scan} 
except the LHC search and relic density constraints. 
Purple, green, and blue points are subsets of grey points representing solutions with relic density larger than, 
within, and lower than the range in Eq.~(\ref{eq:omega}) respectively. These points also satisfy 
the LHC limits described in Section~\ref{sec:scan}.
}
\label{fig2}
\end{figure}

We now turn to look at the phenomenology of the lightest neutralino as DM candidate. In Fig.~\ref{fig2}, we show the neutralino relic density versus to its mass, as resulting from the standard freeze-out mechanism. Grey points fulfil all constraints discussed in Section \ref{sec:scan} but are excluded by the LHC searches.
Colored points satisfy such limits and highlight whether the neutralino LSP is overabundant, underabundant, or its relic density in the range of Eq.~(\ref{eq:omega}). The purple points are clearly excluded by the DM relic density inferred from CMB observations unless some non-standard dilution mechanism is assumed.\footnote{Another option could be considering a scenario with a light axino ($\widetilde{a}$) LSP. In such a case the axino is non-thermally produced through neutralino decays, such that the resulting $\Omega_{\widetilde a}h^2$ is suppressed by a factor $m_{\widetilde a}/m_{\widetilde \chi_1^0}$ with  respect to the neutralino density at freeze out. Nevertheless, in such a scenario, one has to check that the neutralino decay into axino is fast enough not to spoil the successful predictions of 
Big-Bang Nucleosynthesis (BBN). For a review, see \cite{Choi:2013lwa}.}
%
%%%%%%%%%%%%%%%%%%%%%%%%%%%%%%%%%%%%%%%%%%
%%%%%%%%%%%%%%%%%%%%%%%%%%%%%
On the other hand, blue points are phenomenologically viable, although they can not fully account for the observed DM, barring the case that a non-thermal production mechanism is at work. 
If the neutralino is lighter than about 100 GeV, the correct relic density can be achieved only on the $Z$ and $h$ resonances, 
$m_{\widetilde \chi_1^0} \approx m_Z/2$ and $m_{\widetilde \chi_1^0} \approx m_h/2$. We see from the figure that this possibility is already partially excluded by the LHC searches for heavy (Higgsino-like) neutralinos and charginos decaying $WZ$ and the LSP, as discussed in \cite{Calibbi:2014lga,Ahmed:2017jxl}, roughly
giving the bound shown in Eq.~(\ref{lhc-b}). Above 100 GeV, the LEP bounds do not forbid the LSP to be 
mostly Higgsino so that we can have points featuring a substantial DM underabundance. In fact, our naturalness 
requirement in Eq.~(\ref{eqn:deltaEW}) constrains Higgsinos (and hence our neutralino LSP) to be lighter than about 600 GeV, as we can see from the figure, while a pure Higgsino LSP is underproduced unless it is as heavy as about 1.1 TeV, because of its fast annihilation modes into $SU(2)_L$ gauge bosons.
%
%%%%%%%%%%%%%%%%%%%%%%%%%%%%%
\begin{figure}[t!]
\centering
\subfiguretopcaptrue
\subfigure{
\includegraphics[totalheight=5.5cm,width=7.cm]{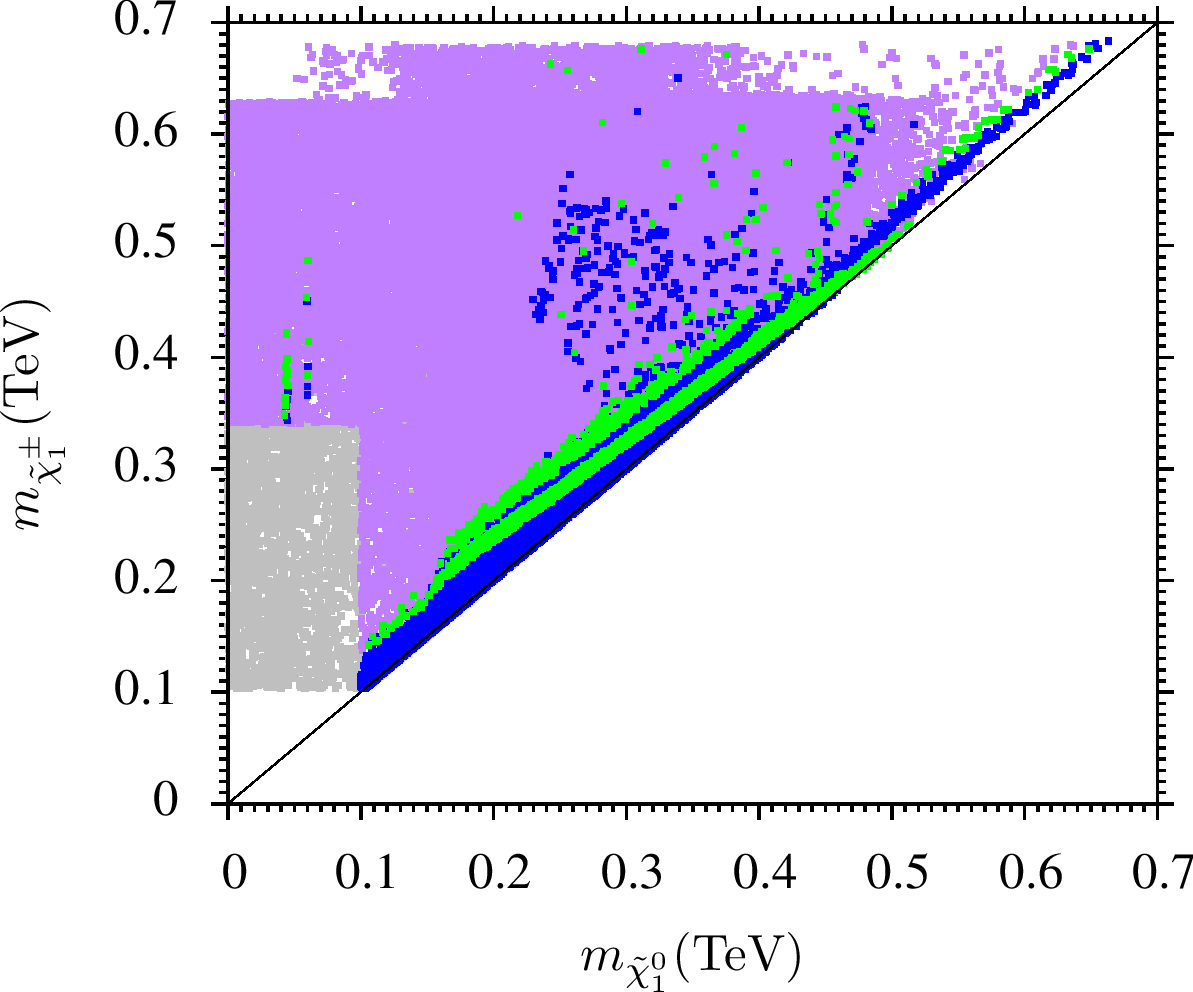}
}
\subfigure{
\includegraphics[totalheight=5.5cm,width=7.cm]{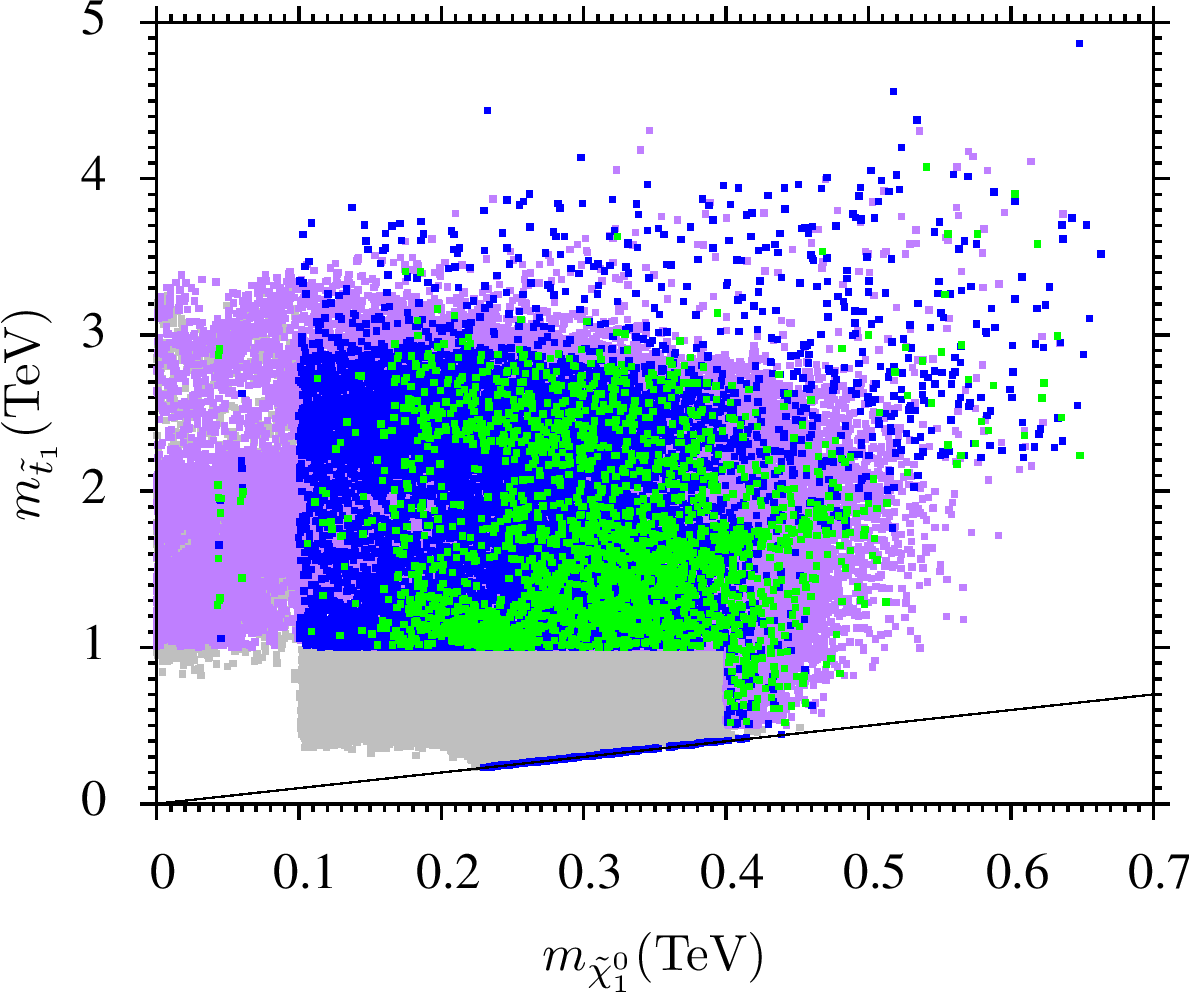}
}
\subfigure{
\includegraphics[totalheight=5.5cm,width=7.cm]{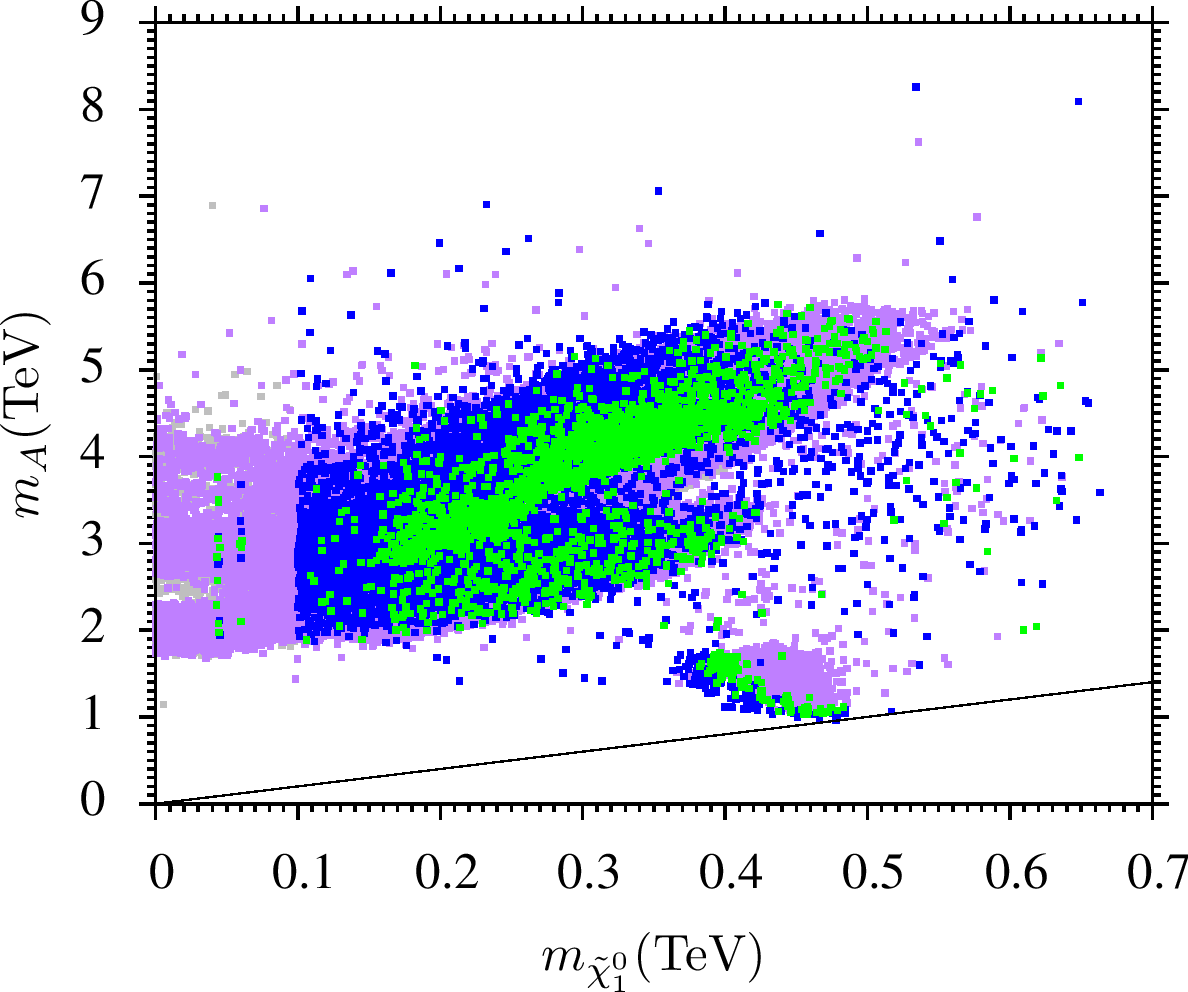}
}
\subfigure{
\includegraphics[totalheight=5.5cm,width=7.cm]{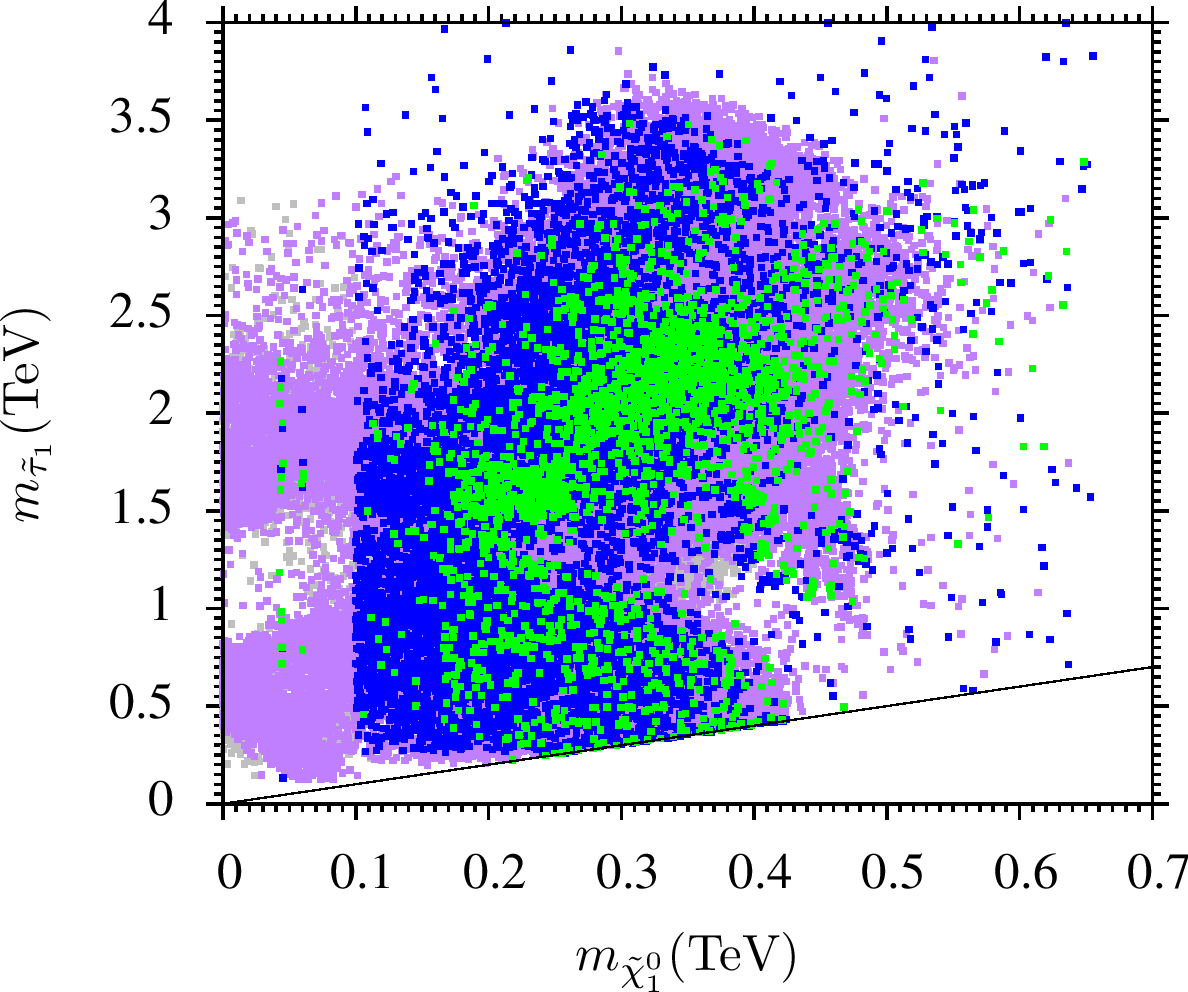}
}
\caption{\small Points of the scan with $\Delta_{EW}\leq$ 100 shown on the plane of the LSP mass $m_{\widetilde \chi_{1}^{0}}$ and the lightest chargino mass $m_{\widetilde \chi_{1}^{\pm}}$ (top left), stop mass $m_{\widetilde t_{1}}$ (top right), CP-odd Higgs mass $m_A$ (bottom left), lightest stau mass $m_{\widetilde \tau_{1}}$ (bottom right). Same color code as in Fig.~\ref{fig2}. }
\label{fig3}
\end{figure}
%%%%%%%%%%%%%%%%%%%%%%%%%

In order to identify the neutralino annihilation or coannihilation mechanisms responsible for the results shown 
in Fig.~\ref{fig2},
we can look at the plots of Fig.~\ref{fig3}, where the same points are displayed in terms of the neutralino mass and the masses of the other particles of the model that are possibly light.
In the top-left panel, we plot the chargino vs.~neutralino mass, from which we can see what already mentioned above: 
below $m_{\widetilde \chi_{1}^{0}}\approx 100$ GeV the relic density constraint in Eq.~(\ref{eq:omega}), 
can be only satisfied at the $Z$ and $h$ resonances,
where a relatively heavy Higgsino (thus chargino) is possible, since the resonant enhancement provides large annihilation rates even for relatively low Higgsino
component in $\chi_{1}^{0}$. We can also see that this possibility is partially excluded by the LHC neutralino-chargino searches giving the approximate bound in Eq.~(\ref{lhc-b}). Above a DM mass of 100 GeV, the underabundant blue points typically correspond to a Higgsino-like neutralino, hence neutralino and chargino are degenerate. Also, most of points with the correct relic density feature $m_{\widetilde \chi_{1}^{\pm}}\gtrsim m_{\widetilde \chi_{1}^{0}} $, which
means a large Bino-Higgsino mixing. As we will see, this possibility is now excluded by direct detection searches. There are however some green points far from the diagonal, corresponding to other annihilation mechanisms, as it is clear from the other plots in Fig.~\ref{fig3}.

In the top-right plot, where we show the stop mass, we can see that neutralino-stop coannihilations are severely constrained by our limits in Eqs.~(\ref{lhc-c}\,-\ref{lhc-e}).
 Apart from a small region with $m_{\widetilde \chi_{1}^{0}}\gtrsim 400$ GeV, the coannihilation strips only survives for a very small mass splitting that gives in turn $\Omega_\chi h^2 \ll 0.12$. The bottom row of the Fig.~\ref{fig3} shows instead that efficient annihilations through a CP-odd Higgs $A$ (bottom left) and coannihilation with the stau (bottom right) are possible in some corners of the parameter space. 
 In particular, in the bottom-right plot we show that $A$ is typically heavy, but there is region where the neutralino mass is
 approaching the resonant condition $m_{\widetilde \chi_1^0} \approx m_A/2$ (a solution often called `$A$-funnel'). Large part of the plane with light $A$ and ${\widetilde \chi_1^0}$ is excluded
 by the interplay of the $B_s \rightarrow \mu^+ \mu^-$ and $b\to s \gamma$ constraints in combination with the Higgs mass requirement (for a discussion see {\it e.g.}~\cite{Calibbi:2011ug}).
%
%%%%%%%%%%%%%%%%%%%%%%%%%%%%%
\begin{figure}[t!]
\centering
\subfiguretopcaptrue
\subfigure{
\includegraphics[totalheight=5.5cm,width=7.cm]{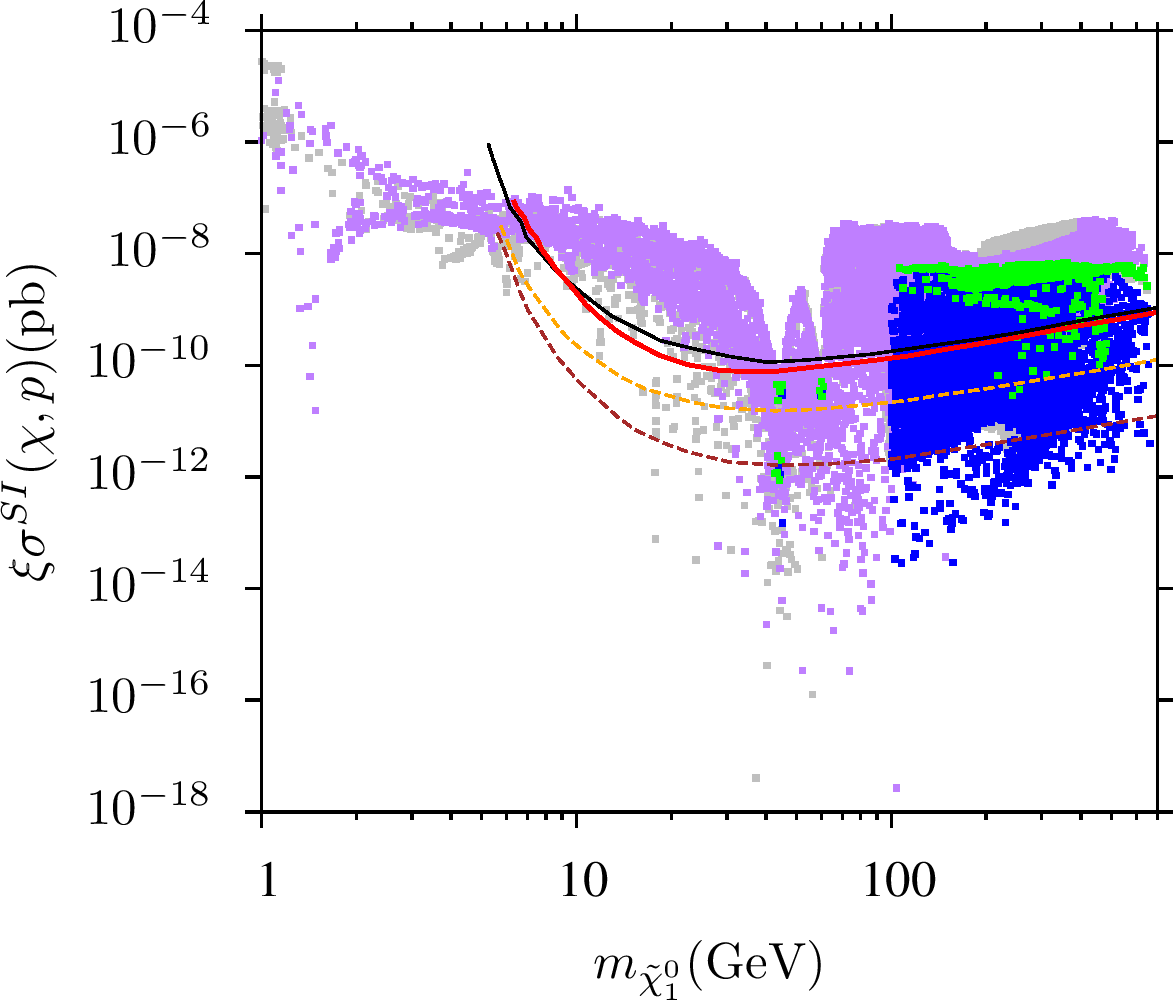}
}
\subfigure{
\includegraphics[totalheight=5.5cm,width=7.cm]{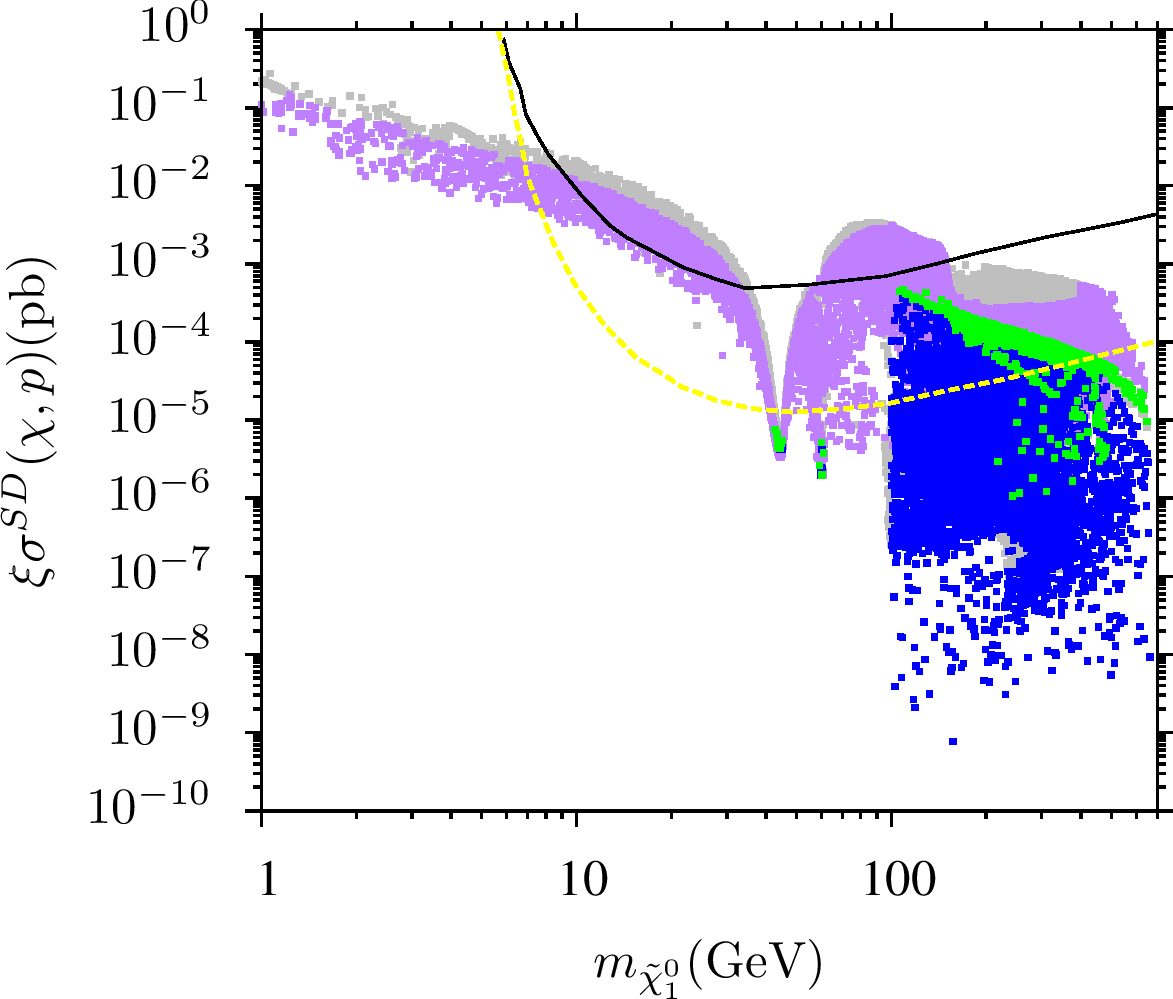}
}
\caption{\small Rescaled spin-independent (left) and spin-dependent (right) neutralino-proton scattering cross section
vs.~the neutralino mass. The scaling factor is defined as $\xi\equiv \Omega_\chi h^2 /0.12$. The color code is the 
same as in Figs.~\ref{fig2} and \ref{fig3}.
In the left plot, the solid black and red lines respectively represent the current LUX \cite{Akerib:2016vxi} and XENON1T \cite{Aprile:2017iyp} bounds, while the dashed orange and brown lines show the projection of future limits \cite{Aprile:2015uzo} of XENON1T with 2 $t\cdot y$ exposure and XENONnT with 20 $t\cdot y$ exposure, respectively. In the right plot, the black solid line is the current LUX bound \cite{Akerib:2017kat} and the yellow dashed line represents the future LZ bound \cite{Akerib:2016lao}.
}
\label{fig4}
\end{figure}
%%%%%%%%%%%%%%%%%%%%%%%%%%%%%

We now consider the impact of the current and future DM searches on our model, still focusing on the `natural' regions of the parameter space as in Eq.~(\ref{eqn:deltaEW}).
In Fig.~\ref{fig4}, we plot the spin-independent (left panel) and the spin-dependent (right panel) neutralino-proton scattering cross sections rescaled by
a factor $\xi =  \Omega_\chi h^2 /0.12$, which accounts for the depletion of the bounds as a consequence of a low local neutralino abundance in the cases that it can not fully account for the observed DM relic density. The present limits from direct detection experiments are shown as solid lines.
As we can see, these bounds strongly affect our parameter space, especially the spin-independent one. While the $h$ and $Z$ resonances are not severely constrained at the moment, most of the (green) points compatible with the observed DM relic density (\ref{eq:omega}) are excluded by the limits recently published by LUX and XENON1T. In particular, this is the case of the configurations with substantial Bino-Higgsino mixing, because this induces a sizable $\widetilde \chi_{1}^{0}-\widetilde \chi_{1}^{0}-h$ coupling. This scenario -- some times referred to as `well-tempered' neutralino \cite{well-tempered1} -- is thus excluded in our D-brane
model. For recent discussions on the direct-detection constraints on well-tempered neutralinos, see also \cite{well-tempered2,well-tempered3}.
The green points that survive the bound correspond to a Bino-like neutralino with the relic density bound fulfilled through a CP-odd Higgs exchange or stau coannihilation, as illustrated in the second row of Fig.~\ref{fig3}. 

The plots in Fig.~\ref{fig4} also show that, interestingly, the future sensitivity of direct searches is capable to test almost completely our D-brane scenario with $\Delta_{\rm EW} < 100$ not only for the neutralino relic density 
in the range of Eq.~(\ref{eq:omega}), but also for most of the blue points with an underabundant neutralino 
due to mainly Higgsino-like LSP (for a discussion of this scenario, we refer to \cite{Badziak:2015qca}). 
In summary, we see that combining our naturalness requirement with relic density constraints (that rule out the purple points) makes
our model tightly constrained and in principle fully testable by DM searches, unless a substantial deviation from the standard thermal freeze-out paradigm is assumed.

%%%%%%%%%%%%%
\begin{table}[t]\hspace{-1.0cm}
\scriptsize{
\centering
\begin{tabular}{|c|ccccccc|}
\hline
\hline
                 & Point 1 & Point 2 & Point 3 & Point 4 & Point 5 & Point 6 & Point 7\\
\hline
$m_{L}$        & 1805 & 2048.4   & 2071.4    & 2167.7   & 5517     & 1559.4 & 4141.3\\
$m_{R}$        & 1839.7 & 1793   & 3038.5    & 1974.9   & 3383.8     & 1463.4  & 2356.7\\
$M_{1} $       & 72.02 & 105.01  & -1065.2   & 267.71   & -1608.6    & -635.52 & -607.34\\
$M_{2}$        & -3090.8 & -3057   & -4888.4 & -3158.2   & 3038.5   & -3237.7 & 3142.1\\
$M_{3}$        & -1473.2 & -1566.7  & -1479.1  & -1489.1 & 1887.7   & -1185.6 & 2305.1\\
$A_0$          & 1045.3 & 959.36  & 3753.5    & 658.41   &-931.66     & 2692.1 & -744.84\\
$\tan\beta$    & 17.3  & 18.8            & 54.6  & 18.3   &44.7        & 12.8   & 46.5\\
$m_{H_u}=m_{H_d}$   & 2397.4  & 2551.8  & 2886.1  & 2519.8 &4682.9  & 2321.6 & 4137.2\\
\hline
$\mu$            &384   & 425        &   638  & 158     &199    & 375  & 88\\
$\Delta_{\rm EW}$ &{\bf 37}  &{\bf 43}  & {\bf 99} &{\bf 32} & {\bf 14}  &{\bf 63}  & 88\\
%$\Delta_{HS}$    & 1406  & 1599        &    2068        & 1533 &1488 &1369 & 4170\\
\hline
\hline
$m_h$            &122   & 122     & 125       & 122  & 122   &126 & 123 \\
$m_H$            & 2972 & 3048    & 1064      &3053  & 3115  &3042 & 2749\\
$m_A$            & 2953 & 3028    & {\bf 1057}       & 3094  &3196  &3022 & 2730\\
$m_{H^{\pm}}$    & 2973 & 3049    & 1069      &3054  & 3116   &3043 & 2750\\
\hline
$m_{\widetilde{\chi}^0_{1,2}}$
                 & {\bf 45}, 396 & {\bf 61}, 438    & {\bf 472}, 653    & {\bf 117}, 166 & {\bf 205 },206 &{\bf 270}, 387 & {\bf 294}, 620\\
$m_{\widetilde{\chi}^0_{3,4}}$
                 & 400, 2546   & 441, 2524 & 656, 4042    & 186, 2606 & 753, 2564 &390, 2688 & 624, 2638\\
$m_{\widetilde{\chi}^{\pm}_{1,2}}$
                 & 375, 2515  & 414, 2491 &  621, 4022    &{\bf 154}, 2572&{\bf 213}, 2531  &372, 2684 & 637, 2603\\
\hline
$m_{\widetilde{g}}$  & 3204 & 3390 &  3255  & 3248 & 4128  &2627 & 4866\\
\hline
$m_{ \widetilde{u}_{L,R}}$
                 & 3743, 3254 & 3965, 3358 & 4509, 4052  & 3965, 3351 & 6411, 4765 & 3344
                 ,2639 & 6085, 4683\\
$m_{\widetilde{t}_{1,2}}$
                 & 1953, 3260 & 1999, 3463  & 1970, 3183  & 1027, 2978 & 2275, 5308 & {\bf 272}, 2774 & 2657,5115\\
\hline $m_{ \widetilde{d}_{L,R}}$
                 & 3744, 3257 & 3966, 3361  & 4510, 4051  & 3966, 3354 & 64112, 4753 & 3345, 2639 &6085, 4682\\
$m_{\widetilde{b}_{1,2}}$
                 & 3156, 3270 & 3241, 3476  & 2695, 3193  & 3244, 3486 & 3788, 5354 & 2536, 2818 &3785, 5145 \\
\hline
$m_{\widetilde{\nu}_{1,2}}$
                 & 2656 & 2810  & 3711  & 2944 & 5519 &2576 & 4578\\
$m_{\widetilde{\nu}_{3}}$
                 &  2625 & 2773 & 3213  & 2911& 5170 &2565 & 4269\\
\hline
$m_{ \widetilde{e}_{L,R}}$
                & 2657, 1838 & 2811, 1790 & 3710, 3064  & 2944, 1975 & 5514, 3425 & 2573, 1479 & 4576,2356\\
$m_{\widetilde{\tau}_{1,2}}$
                & 1746, 2626 & 1673, 2774 &  1493, 3200 & 1870, 2910 & 2059, 5164 &1386, 2555 & {\bf 299},4265\\
\hline
$\sigma_{\rm SI}({\rm pb})$
                & $4.44\times 10^{-11}$ & $4.20\times 10^{-11} $ & $ 1.48\times 10^{-10} $ & $ 5.38\times 10^{-9}$&${\bf 1.05\times 10^{-11}}$ & ${\bf 7.42\times 10^{-10}} $ & ${\bf 7.44\times 10^{-11}} $\\

$\sigma_{\rm SD}({\rm pb})$
                & $5.53\times 10^{-6}$ & $3.79\times 10^{-6}$ & $3.45\times 10^{-6}$ & $3.48\times 10^{-4} $&${3.49\times 10^{-7}}$ & ${\bf 2.17\times 10^{-5}} $ & ${\bf 1.55\times 10^{-6}} $\\

$\Omega_{\chi}h^{2}$&  0.104 & 0.110   & 0.101  & 0.129& {\bf 0.007 } & {\bf 0.002} & {\bf 0.128}\\
\hline
\hline
\end{tabular}
\caption{The particle spectra and properties of neutralino DM for a set of representative points for
 different regions of the viable parameter space. See the text in Section \ref{sec:conclusions} for details. The first block shows high-energy parameters defined at the GUT scale, while the others contain low-energy quantities.
All quantities with mass dimension $[M]$ are in the unit of GeV.
}
}
\label{table1}
\end{table}
%%%%%%%%%%%%%%%%%%%%%%%%%%%

%%%%%%%%%%%%%%%%%%%%%%%%%%%%%%%%%%%%%%%%%%%%

%%%%%%%%%%%
\section{Summary and Conclusions}
\label{sec:conclusions}

We have revisited the predicted low-energy spectra of SUSY particles in a realistic D-brane model
with a particular focus on the recent LHC and DM constraints
in the regions of the parameter space characterized by low levels of fine-tuning. Relatively natural solutions are possible due to the generically non-universal gaugino mass terms predicted by our model at the GUT scale, cf.~the boundary conditions (\ref{ssb}).
In our phenomenological survey presented in Section \ref{sec:numerics}, we have found that, although several (co)annihilation modes can account for the DM abundance inferred from CMB observations, experimental constraints, 
in particular the LHC searches and direct DM detection, set very severe bounds on the parameter space. Interestingly, next generation direct detection experiments should be able to test the low tuning configurations of the model, as a consequence of the upper bound on the LSP mass ($\lesssim 600$ GeV) set by such requiring a tuning not worse than the percent level.

We summarize our findings by showing in Table \ref{table1} some points of the parameter space representative of the different regions identified and discussed in the previous section. All points feature a rather heavy spectrum and/or small mass splittings that make them not easily accessible at the LHC with the possible exception of Higgsino sector.
Points 1, 2, and 3 respectively represent the $Z$-resonance, Higgs-resonance, and $A$-funnel solutions, resulting in a neutralino relic density in the range quoted in Eq.~(\ref{eq:omega}). In all three cases, the spin-independent neutralino-nucleon cross section is such that the current bounds are evaded but a signal
at currently running or future direct detection experiments is expected.
Point 4 is an example of the Bino-Higgsino mixed dark matter, which is already excluded by direct detection, 
while point 5 features a (light) mostly-Higgsino LSP, so that
it is still viable because of the suppressed relic abundance. Despite that, point 5 exemplifies solutions with underabundant neutralinos in the reach of direct detection experiments, as discussed in section \ref{sec:numerics}. 
Points 6 and 7 respectively represent solutions with efficient stop- and stau-neutralino coannihilations. In the former case, the small stop-neutralino mass splitting gives as a result underabundant neutralino DM. Again, both scenarios predict
a scattering cross section at levels observable at direct detection experiments.

%%%%%%%%%%%%%%%%%%%
\section*{Acknowledgements}
TL was supported in part by the Projects 11475238 and 11647601 supported by National Natural Science Foundation 
of China, and by Key Research Program of Frontier Science, CAS. WA was supported by the CAS-TWAS Presidents 
Fellowship Programme. The numerical results described in this paper have been obtained via the HPC Cluster 
of ITP-CAS. SR would like to thank TL for warm hospitality and the Institute of Theoretical Physics, 
CAS, P. R. China for providing conducive atmosphere for research where part of this work has been carried out.

%%%%%%%%%%%%%%%%%%%%%%%%%%%%%
%%%%%%%%%%%%%%%%%%%%%%%%%%%%%%%%%%%%%%%%%%%
%%%%%%%%%%%%%%%%%%%%%%%%%%%%%%%%%%%%%%%%%%%%%%%%
%%%%%%%%%%%%%%%%%%%%%%%%%%%%%%%%%%%%%%%
%\newpage
%%%%%%%%%%%%%%%%%%%%%%%%%%%%%%%%%

\end{document}